\documentclass[reprint,floatfix,superscriptaddress,showpacs,aps,pra,tightenlines]{revtex4-1}

\usepackage{epsfig}
\usepackage{amsmath}
\usepackage{color}
\usepackage{multirow}
\usepackage{times}

\newcommand{\SWAP}{\textsc{swap}}
\newcommand{\CZ}{\textsc{cz}}
\newcommand{\CNOT}{\textsc{cnot}}

\begin{document}

\bibliographystyle{apsrev}

\title{Hybrid-system approach to fault-tolerant quantum communication}

\author{Ashley M. Stephens}\email{astephens@unimelb.edu.au}
\affiliation{National Institute for Informatics, 2-1-2 Hitotsubashi, Chiyoda-ku, Tokyo 101-8430, Japan}

\author{Jingjing Huang}
\affiliation{NTT Basic Research Laboratories, 3-1 Morinosato-Wakamiya, Atsugi, Kanagawa 243-0198, Japan}
\affiliation{California Institute of Technology, Pasadena, California 91225, USA}

\author{Kae Nemoto}
\affiliation{National Institute for Informatics, 2-1-2 Hitotsubashi, Chiyoda-ku, Tokyo 101-8430, Japan}

\author{William J. Munro}
\affiliation{NTT Basic Research Laboratories, 3-1 Morinosato-Wakamiya, Atsugi, Kanagawa 243-0198, Japan}
\affiliation{National Institute for Informatics, 2-1-2 Hitotsubashi, Chiyoda-ku, Tokyo 101-8430, Japan}

\date{\today}

\frenchspacing

\begin{abstract}
We present a layered hybrid-system approach to quantum communication that involves the distribution of a topological cluster state throughout a quantum network. Photon loss and other errors are suppressed by optical multiplexing and entanglement purification. The scheme is scalable to large distances, achieving an end-to-end rate of 1 kHz with around 50 qubits per node. We suggest a potentially suitable implementation of an individual node composed of erbium spins (single atom or ensemble) coupled via flux qubits to a microwave resonator, allowing for deterministic local gates, stable quantum memories, and emission of photons in the telecom regime.
\end{abstract}
 
\maketitle

\section{Introduction}

Quantum communication is predicated on the ability to quickly and reliably entangle two or more quantum systems that are separated by geographically large distances \cite{Gisin1}. What makes quantum communication difficult is that photons are easily lost during transmission, due to attenuation in optical fibers and inefficient optical coupling. Photon loss, in addition to errors due to decoherence and imprecise quantum control, must be overcome before quantum communication can be a useful technology.

One proposal for reliable quantum communication is to use a network of relatively simple devices known as quantum repeaters, analogous to how optical amplifiers are used in classical communication \cite{sangouard10}. Rather than attempting to directly entangle a pair of qubits at the ends of the communication channel, the channel is divided into short segments and pairs of qubits in adjacent nodes are entangled and purified to sufficiently high fidelity \cite{Bennett2,Deutsch1}. Then, the range of entanglement is extended to the endmost nodes \cite{Bennett1}. The basic elements of such a scheme have been demonstrated in the laboratory \cite{sangouard10,pan1,yuan1}. However, in practice, the length of the communication channel will be limited by the coherence time of the quantum memories, as the time required to purify entanglement increases with distance \cite{Hartmann1}.

As experiments begin to scale beyond the laboratory, attention should turn to proposals for quantum communication that account for all sources of error. Here, we present a scheme that combines aspects of the orthodox repeater network with fault-tolerant error correction. The foundation of the scheme is optical multiplexing, which serves to reduce the effective loss rate between nodes \cite{Munro1}. Then, purification is used to increase the fidelity of entanglement between nodes \cite{Hartmann1}. Finally, this entanglement is used to generate a three-dimensional topological cluster state \cite{Raussendorf3}. It is using this state that we can ensure that communication of logical qubits over large distances is reliable. Topological cluster-state error correction requires low connectivity---desirable in the context of network communication---but exhibits a relatively high threshold \cite{Barrett1}. We find the scheme is able to tolerate photon loss of 80\%, communication channel errors of 10\% per qubit, and local gate errors of 0.1\% per gate. Under these conditions, the scheme is scalable to large distances, achieving an end-to-end rate of approximately 1 kHz with around 50 qubits per node.  

Quantum communication of logical qubits protected by error correction was proposed by Jiang {\it et al.} \cite{jiang09}. Other authors have proposed schemes based on graph states \cite{Perseguers1,Perseguers2}, parity codes \cite{munro12}, and the surface code \cite{Fowler1}. Very recently, Li {\it et al.} proposed a scheme for quantum communication using a topological cluster state \cite{Li3}. The distinguishing feature of their proposal is that entangling operations within the repeater nodes are allowed to fail. For all of these proposals, it will be important to identify physical systems that meet the requirements of scalability, to establish threshold error rates for the fundamental components, and to understand the associated performance and resource requirements. For our proposal, we suggest a potentially suitable implementation of an individual node composed of erbium spins (single atom or ensemble) coupled via flux qubits to a microwave resonator, allowing for deterministic local gates, stable quantum memories, and emission of photons in the telecom regime. 

This paper is organized as follows: in Sec.~\ref{ED} we describe how to reliably distribute entanglement between nodes in the repeater network, in Sec.~\ref{EP} we describe how to purify the entanglement to sufficient fidelity and then use this entanglement to create a topological cluster state throughout the network, in Sec.~\ref{PO} we determine the performance of the scheme and the associated overhead, and in Sec.~\ref{PSR} we outline a potential implementation of the scheme.

\begin{figure}[htb]
\begin{center}
\resizebox{70mm}{!}{\includegraphics{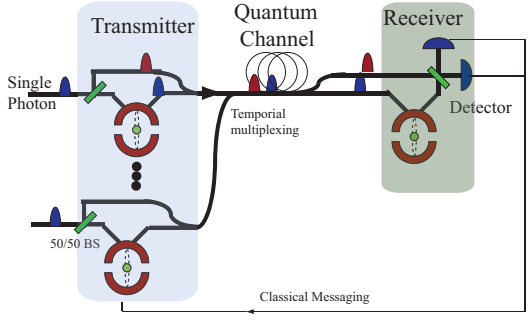}}
\end{center} 
\vspace*{-10pt}
\caption{Scheme for distributing entangled pairs between adjacent repeater nodes using a multiplexed transmitter-receiver model \cite{Munro1}.}
\label{fig:multiplex}
\end{figure}

\section{Entanglement distribution} 
\label{ED}

Our starting point is a quantum network in which we can generate entanglement between a pair of qubits in neighboring nodes, each being inside an optical cavity, using a multiplexed transmitter-receiver model \cite{Munro1}. The scheme is depicted in Fig.~\ref{fig:multiplex}. Before considering this scheme, however, we will outline a more simple version.

First we prepare a matter-based qubit at each of the transmitting and receiving nodes in an equal superposition of its ground and first excited states, $( |0\rangle_q+|1\rangle_q )/ {\sqrt 2} $. We also prepare a single photon at the transmitting node, $|1\rangle_p$. This gives us an initial state $( |0\rangle_{q1}+|1\rangle_{q1} ) |1\rangle_p ( |0\rangle_{q2}+|1\rangle_{q2} )/2$. We then split the photon on a 50:50 beamsplitter to create the two-mode state $ ( |1\rangle_{p1} |0\rangle_{p2}+|0\rangle_{p1} |1\rangle_{p2} ) / {\sqrt 2}$. When the second mode interacts with the qubit in the first optical cavity, it picks up a $\pi$ phase shift if and only if the qubit is in the state $|1\rangle_{q1}$ and a photon is present in the second mode \cite{DIT,DIT2}. After this interaction our system has evolved to 
\begin{eqnarray}
& & \frac{1}{2\sqrt 2}  |0\rangle_{q1} ( |1\rangle_{p1} |0\rangle_{p2}+|0\rangle_{p1} |1\rangle_{p2} ) ( |0\rangle_{q2}+|1\rangle_{q2} ) \nonumber \\
&+& \frac{1}{2\sqrt 2} |1\rangle_{q1} ( |1\rangle_{p1} |0\rangle_{p2}-|0\rangle_{p1} |1\rangle_{p2} ) ( |0\rangle_{q2}+|1\rangle_{q2} )
\end{eqnarray}
The two-mode photon field is then temporally multiplexed and transmitted to the receiving node. Upon arrival, the multiplexing is reversed and the second mode interacts with the qubit in the second optical cavity, again picking up a $\pi$ phase shift if and only if the qubit is in the state $|1\rangle_{q2}$ and a photon is present in the second mode. We then recombine the photonic modes on a 50:50 beamsplitter, so the state of the system is
\begin{eqnarray}
\frac{1}{2} |0\rangle_{q1} |1\rangle_{p1} |0\rangle_{p2} |0 \rangle_{q2}&+& \frac{1}{2} |1\rangle_{q1} |1\rangle_{p1} |0\rangle_{p2} |1 \rangle_{q2} \nonumber \\
+\frac{1}{2} |0\rangle_{q1} |0\rangle_{p1} |1\rangle_{p2} |1 \rangle_{q2}&+& \frac{1}{2} |1\rangle_{q1} |0\rangle_{p1} |1\rangle_{p2} |0 \rangle_{q2} 
\end{eqnarray}
We finish by measuring the presence or absence of the photon at either output port. Detection of a photon in the $p1$ mode implies we have created the two-qubit state 
$( |0\rangle_{q1}  |0 \rangle_{q2}+ |1\rangle_{q1} |1 \rangle_{q2} ) / {\sqrt 2} $, while detection of a photon in the $p2$ mode implies the state $( |0\rangle_{q1}  |1 \rangle_{q2}+ |1\rangle_{q1} |0 \rangle_{q2} ) / {\sqrt 2} $. A third outcome is possible---no photon will be detected if it has been lost, either in its interaction with the matter qubits or during transmission between nodes, or if the detector fails. In this case we discard the attempt. In any case, a classical signal is sent to the transmitting node indicating whether the attempt has been successful or not. 

In general, the probability of successfully generating a highly entangled state is
\begin{equation}
p_{\rm tot} = p_{\rm single} \times p_{\rm coupling}^2 \times p_{\rm detector} \times e^{-{\cal L/L}_0},
\end{equation}
 where $p_{\rm single}$ is the probability of generating a single photon, $ p_{\rm coupling}$ is the probability of successfully coupling the single photon into and out of the optical  cavity, $p_{\rm detector}$ is the efficiency of the detector, ${\cal L}$ is the distance between the two nodes, and  ${\cal L}_0$ is the attenuation length of the fiber (approximately 25 km for commercial telecom fiber). In practice, $p_{\rm tot}$ will be much less than one---for ${\cal L}=10$ km, $p_{\rm tot}=0.2$ is optimistic \cite{optimistic}. 

We have assumed that the photon field will pick up a $\pi$ phase shift if and only if the qubit is in the state $|1\rangle_{q1}$ and a photon is present in the field. This is a stringent requirement, which can be relaxed. Instead of a $\pi$ phase shift, we can consider the situation where a $\theta$ phase shift is achieved. Now only one detection event (detection of a photon in the $p2$ mode) heralds the generation of a suitably entangled state. The probability of success is reduced by $(1-\cos \theta)/4$.

We now turn to the multiplexed schemed depicted in Fig.~\ref{fig:multiplex}. In this scheme, we begin by preparing several qubits in the transmitting node in the state $(|0\rangle+|1\rangle)/\sqrt{2}$, in addition to the qubit in the receiving node. In the transmitting node, each qubit interacts (and becomes entangled) with its own individual single photon. These photons are temporally multiplexed and transmitted to the receiving node, preceded by a classical herald. When the herald is received, the first photon is coupled with the qubit at that node. The subsequent detection of the photon heralds the generation of an entangled state between the first qubit in the transmitting node and the qubit in the receiving node. In practice however, it is likely the photon will have been lost. In this case, the qubit in the receiving node is re-prepared for the second photon in the multiplexed signal and the sequence of entanglement and measurement is repeated. This is done until a success is reported.  Once that qubit is entangled, the remaining photons are sent to another qubit in the same node, and so on until all incoming photons have been depleted and a number of entangled pairs have been generated. To further improve the efficiency of the scheme, we can transmit qubits from both nodes simultaneously, leaving a small number of qubits on each side as receivers.

It is important that multiplexing be tolerant of the various errors that can arise. By design, photon loss is tolerated \cite{sangouard10,collins07,Simon1,Munro1}. An error during preparation of the matter-based qubits or an error due to decoherence of quantum memories during transmission of the photons will lower the fidelity of the entanglement in the event that the attempt succeeds. Such an error cannot affect more than one pair of qubits that is eventually accepted, so, for these errors, attempts are effectively independent. Detection errors might not be so benign. If an attempt to generate entanglement is successful but the photon is not detected because the detector failed, then the pair of qubits is rejected as if the photon was lost. However, in the case of a dark count, we might accept a pair of qubits that are not, in fact, entangled. Thus, we require single-photon detectors that are reliable enough so that the cumulative probability of a dark count during the series of attempts does not limit the fidelity of the entanglement. Recent results suggest that this can be achieved \cite{Dorenbos}. Ultimately, we are left with a number of entangled pairs of some fidelity, $F$, shared between adjacent nodes in the network.

\section{Purification and \\ topological error correction} 
\label{EP}

Next, to increase the fidelity of entanglement between adjacent nodes, we turn to purification \cite{Bennett2,Deutsch1}. Each round of purification requires classical communication between nodes. Already purified pairs can be used in successive rounds of purification until sufficient fidelity is achieved, following various strategies \cite{Meter1}. However, typical purification schemes are slow and require many rounds of classical communication. Instead, we will consider a variation of purification based on Calderbank-Shor-Steane error-correction codes, which generally requires fewer rounds \cite{Aschauer1,Hartmann1}.

\begin{figure}[t]
\begin{center}
\resizebox{87mm}{!}{\includegraphics{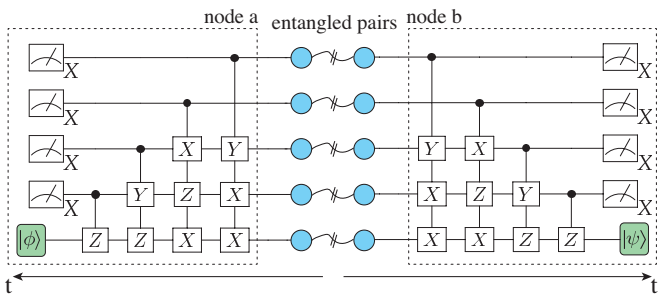}}
\end{center} 
\vspace*{-14pt}
\caption{Purification circuit based on an encoding circuit for the [[5,1,3]] code. The circuit is executed independently at two adjacent nodes, with the input being five entangled pairs (in blue) shared between those nodes. Then, the measurement results from both nodes are combined and the remaining two qubits (one at each node, in green) form the output pair. This is equivalent to a standard encoding-transmission-decoding scheme \cite{Aschauer1}. The circuit is not fault tolerant, but residual errors will be corrected with topological error correction.} 
\label{fig:circuit}
\end{figure}

\begin{figure}[t]
\begin{center}
\resizebox{85mm}{!}{\includegraphics{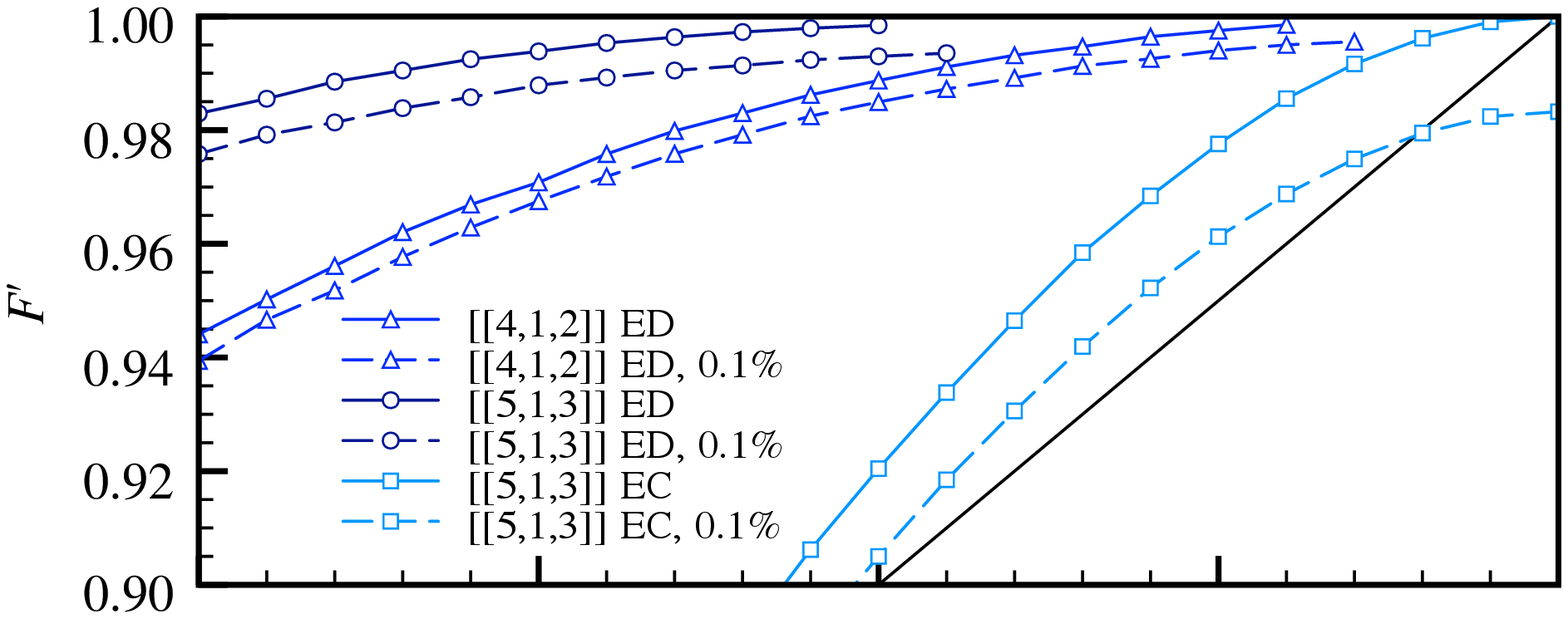}}
\resizebox{85mm}{!}{\includegraphics{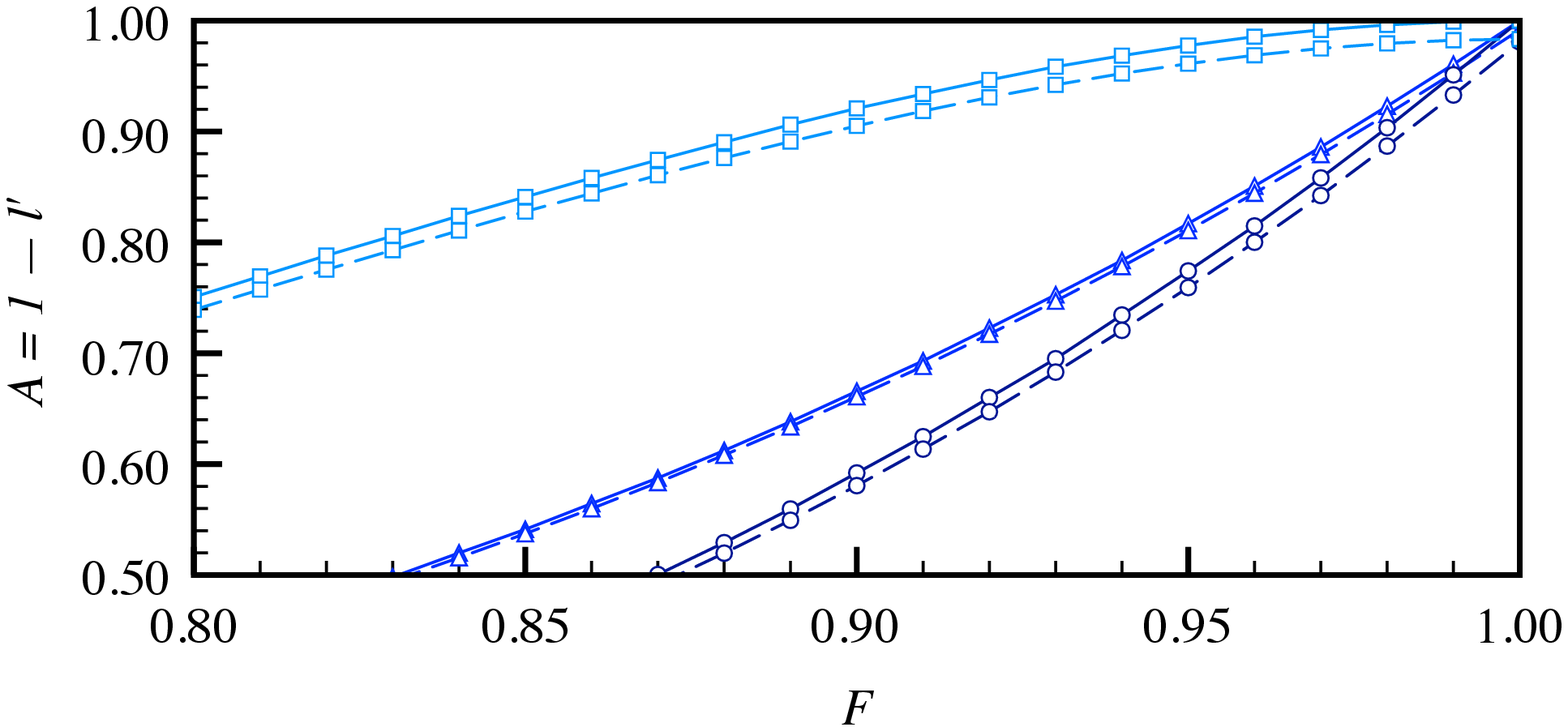}}
\end{center} 
\vspace*{-15pt}
\caption{Results of numerical simulation of purification in error-detection (ED) and error-correction (EC) modes. The output fidelity, $F^\prime$, and acceptance probability, $A=1-l^\prime$, depend on the input fidelity, $F$, where $l^\prime$ is the probability that the output pair is rejected. Solid lines are for perfect local gates and dashed lines are for local gates with an error rate of 0.1\%.}
\label{fig:purification}
\end{figure}

For an $[[n,1,d]]$ quantum code, where $d$ is the code distance, $n$ entangled pairs are required for each round. A simple $n$-qubit circuit (related to an encoding circuit of the code) is executed at both nodes and $n-1$ qubits are measured, leaving one output pair \cite{Aschauer1}. See Fig.~\ref{fig:circuit} for an example. The implementation of the local gates required to perform this circuit will be discussed in Sec.~\ref{PSR}. The measurement outcomes can be used to infer the presence or absence of errors, leading to two possible modes of operation. In either mode, if no errors are detected we keep the output pair. Then, whenever an error is detected, we can reject the output pair (error-correction mode). Alternatively, if $d>2$, for some errors we can apply a correction to the output pair (error-correction mode).

Figure \ref{fig:purification} shows the performance of two codes in both modes. The fidelity of the output pair, $F^\prime$, and the probability that the output pair is accepted, $A$, depend on the input fidelity, $F$. At this point, we have made the assumption that the error rate of the local gates is 0.1\%. For $F=0.900$, two rounds of the [[4,1,2]] code results in $F^\prime=0.997$, while one round of the [[5,1,3]] code results in $F^\prime=0.993$. This is a much greater increase in fidelity per round than for standard two-qubit purification, so fewer rounds (and less classical communication time) will be required. 

\begin{figure} [t]
\begin{center}
\vspace*{5pt}
\resizebox{65mm}{!}{\includegraphics{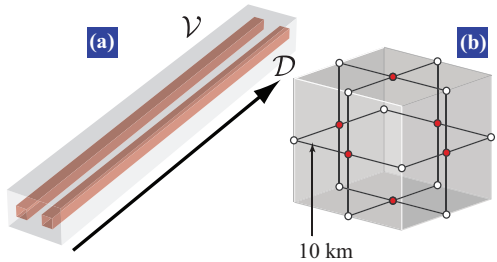}}
\end{center} 
\vspace*{-15pt}
\caption{(a) Communication involves transmitting logical qubits, $\mathcal{D}$, through a topological cluster state. Qubits in the region $\mathcal{V}$ are measured to enable error correction. (b) The connectivity of the network mimics the structure of the topological cluster state. There is one cluster-state qubit per node, and gates between qubits in adjacent nodes are performed using purified entangled pairs.}
\label{fig:cluster}
\end{figure}

With high-fidelity entanglement between adjacent nodes, our goal is to efficiently establish entanglement of arbitrary fidelity that spans the entire network. To do this, we use fault-tolerant error correction. Of the many schemes for error correction, topological cluster-state error correction is particularly promising as it tolerates a relatively high rate of physical errors \cite{Raussendorf3}. The main ingredient of the scheme is the topological cluster state shown in Fig.~\ref{fig:cluster}. One way to prepare such a state is to initialize each qubit in the $|+\rangle$ state then to apply a series of $\CZ$ (controlled-$Z$) gates between neighboring qubits \cite{Raussendorf3}. In our case, the topological cluster state is prepared with one qubit per node. $\CZ$ gates are executed using entangled pairs shared between adjacent nodes \cite{Gottesman1}.

Once the cluster state is prepared, communication proceeds with a sequence of single-qubit measurements. The state is divided into regions that determine the appropriate measurement basis. Logical qubits are defined by regions of the state measured in the $Z$ basis. It is these logical qubits that are transmitted from one end of the network to the other---see Fig.~\ref{fig:cluster}. The rest of the qubits are measured in the $X$ basis to obtain a so-called syndrome of errors. Error correction involves finding the most likely set of errors that is consistent with this syndrome \cite{Raussendorf3}. For error rates below a certain threshold value, increasing the distance of the code by increasing the extent of the topological cluster state will decrease the likelihood of errors affecting the logical qubits.

Since the cluster state is distributed throughout the network, the $\CZ$ gates rely on the availability of entangled pairs. Errors in these gates can either be heralded or unknown. An error is heralded when there is no entangled pair available to perform a $\CZ$ gate. This might be because the output pair of the final round of purification was rejected, which occurs with probability $l^\prime$ (see Fig.~\ref{fig:purification}). In this case, we can chose to abandon the $\CZ$ gate and treat the associated qubits as if they were lost. Then, the lattice is deformed to avoid the lost qubits and error correction of the remaining unknown errors proceeds \cite{Barrett1}. This introduces a trade-off between the fraction of $\CZ$ gates that are abandoned and the threshold error rate of the remaining gates \cite{Barrett1}.

\begin{figure} [t]
\begin{center}
\vspace*{-8pt}
\resizebox{85mm}{!}{\includegraphics{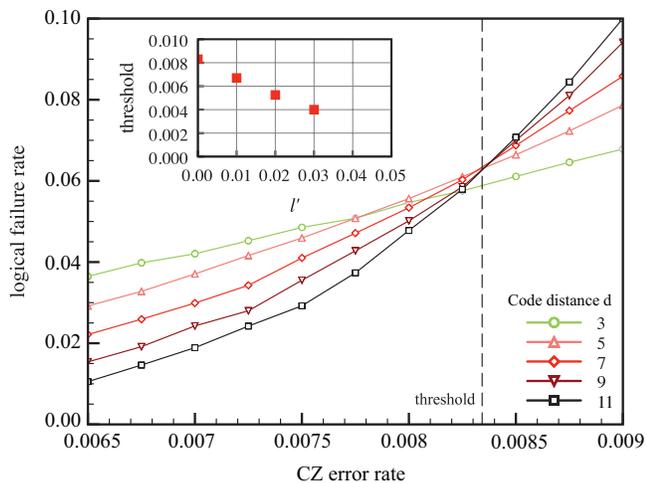}}
\end{center} 
\vspace*{-10pt}
\caption{Results of numerical simulation of topological error correction for various code distances, $d$. A lattice of $d\times d\times d$ unit cells is associated with a distance-$d$ code that may fail for $\geq\lceil d/2\rceil$ errors. The inset shows the threshold as a function of the fraction of abandoned $\CZ$ gates during preparation, $l^\prime$.}
\label{fig:threshold}
\end{figure}

We simulate a topological cluster state with periodic boundary conditions in all three dimensions \cite{sim}. Figure \ref{fig:threshold} shows the logical failure rate as a function of the $\CZ$ error rate for various code distances, where we assume that the error rate of preparation and measurement of qubits at the nodes is 0.1\% and that $\CZ$ gates between nodes never have to be abandoned. We observe a threshold error rate of approximately 0.83\%. Then, the inset to Fig.~\ref{fig:threshold} shows the threshold as a function of the fraction of gates that have to be abandoned, showing the expected tradeoff.

\section{Performance and overhead}
\label{PO}

To determine the performance of our communication scheme we need to consider the combination of entanglement distribution, purification, and error correction. The threshold for topological error correction sets the target for the error rate of the $\CZ$ gates between nodes. In turn, this error rate sets the target for the fidelity of the output pairs of the final round of purification, after accounting for the fact that entangled pairs will not always be available. If these conditions are met, then we can decrease the logical failure rate arbitrarily by increasing the extent of the cluster state, thereby allowing reliable communication over arbitrarily large distances. 

How best to allocate resources to entanglement distribution and purification depends on physical parameters such as the distance between nodes, the rate of photon loss, and the accuracy of the local gates. Here, we assume that our nodes are separated by $L=$10 km (corresponding to a round-trip time of flight of $T_R=0.1$ ms) and that the probability of successfully establishing a raw, entangled pair between two adjacent nodes (before accounting for our multiplexing scheme) is $20\%$ \cite{spacing note}. This efficiency is beyond present experimental capabilities, but will be useful to illustrate our scheme. 

Table \ref{tab:per} outlines a number of strategies to generate entangled pairs that meet or exceed the target fidelity, for various values of the initial fidelity, $F$, and the error rate of the local gates, $p_\textrm{local}$. As an example, assuming that $F=0.907$ and $p_\textrm{local}=0.0008$, with $q=16$ ($q=64$) qubits per node, $N_R=18$ ($N_R=3$) multiples of the round-trip time are required to generate entanglement of sufficient fidelity using one round of the [[5,1,3]] code in error-detection mode. The rate per second at which pairs are generated between adjacent nodes is $R= 1/ (N_R T_R) = 10^4/N_R$. For $q=16$ ($q=64$), $R\sim0.5$ kHz ($R\sim3.3$ kHz). In general, there is a clear tradeoff between $R$ and $q$. Without additional qubits, schemes that require more than one successful round of purification achieve a significantly lower rate, and more accurate local gates mean that fewer successful rounds of purification are required. Our calculations assume worst-case behavior, and it is likely that $R$ could be optimized with more careful scheduling. Ultimately, the communication rate is limited by the time to prepare the cluster state, which is $\sim4T_R$. Quantum memories must be stable over this time, but this requirement is independent of the total communication distance---the extent of the cluster state does not affect its preparation time.

\begin{table}[t]
\begin{center}
\vspace*{4pt}   
\begin{tabular}{ccccccccc}
\hline \hline
$F$					&	0.835		&	0.872			&	0.907			&	0.919			&	0.930		&	0.951		&	0.963 \\
$p_\textrm{local}$		&	0.0005		&	0.0005			&	0.0008			&	0.0005			&	0.0005		&	0.001		&	0.0005 \\
\hline
$l^\prime$			&	0.02			&	0.008			&	0.008			&	0.02				&	0.008		&	0.008		&	0.02 \\
\hline
$q=5$				&	---			&	$154^\dagger$		&	$132^\dagger$		&	$72^\dagger$		& 	$88^{*}$		&	$66^{*}$		&	$54^{*}$ \\
$q=16$				&	$54^{**}$		&	$21^\dagger$		&	$18^\dagger$		&	$12^\dagger$		& 	$8^{*}$		&	$6^{*}$		&	$6^{*}$ \\
$q=32$				&	$36^{**}$		&	$7^\dagger$		&	$6^\dagger$		&	$3^\dagger$		&	$4^{*}$		&	$3^{*}$		&	$3^{*}$ \\
$q=64$				&	$18^{**}$		&	$4^\dagger$		&	$3^\dagger$		&	$2^\dagger$		&	$2^{*}$		&	$2^{*}$		&	$1^{*}$ \\
\hline \hline
\end{tabular}
\caption{Number of round-trip times, $N_\textrm{R}$, required to generate an entangled pair between adjacent nodes with the target fidelity, given initial entangled pairs of fidelity $F$ generated with probability 0.2. Each node contains $q$ matter qubits able to be coupled with deterministic local gates with error rate $p_\textrm{local}$, and $l^\prime$ is the maximum allowed probability that the output pair of the final round of purification is rejected. $^{*}$ indicates that one round of the [[4,1,2]] code is used,  $^{**}$ indicates that two rounds of the [[4,1,2]] code are used, and $^\dagger$ indicates that one round of the [[5,1,3]] code is used. Both codes are used in error-detection mode.}
\label{tab:per} 
\end{center}
\end{table} 

\section{Hybrid-system approach}
\label{PSR}

In Sec.~\ref{ED} we outlined a transmitter-receiver model, in which a matter-based qubit with an optical transition (in the telecom band) is placed in an optical  cavity coupled to an optical fiber. In addition, our scheme requires quantum memories that are stable over multiples of the two-way time of flight between adjacent nodes and near-deterministic local gates. Generally, it is hard to find a single system that satisfies all of these requirements. For instance, a number of atomic systems have optical transitions in the telecom band, but performing local gates between atomic systems in separate cavities is difficult. On the other hand, superconducting circuits can be easily coupled together---spatially separated gap-tuneable flux qubits \cite{Paauw} can be coupled together via a microwave resonator \cite{Fedorov}. However, superconducting circuits do not possess an optical transition. 

A natural solution to this problem is to consider a hybrid-system approach, where superconducting circuits are coupled with atomic-like systems \cite{franco}, enabling local gates and allowing access to an optical transition. Recently it has been shown that information can be transferred between a superconducting circuit and an ensemble of negatively charged nitrogen-vacancy (NV$^-$) centers in diamond \cite{Zhu1,kubo}. The NV$^-$ ensemble couples to the superconducting circuit near 2.88 GHz and has an optical transition, but at 637 nm instead of the telecom band. However, coupling between an ensemble of Er$^{3+}$ spins doped in a Y$_2$SiO$_5$ crystal and a microwave resonator has been demonstrated \cite{Bushev1,Staudt1}. Additionally, Er$^{3+}$ has an $^{4}I_{I5/2} - ^{4}I_{I3/2}$ transition in the telecom $C$ band at $\sim$1540 nm \cite{Lauritzen1} . 

\subsection{Local gates within a repeater node}

\begin{figure} [t]
\begin{center}
\vspace*{5pt}
\resizebox{80mm}{!}{\includegraphics{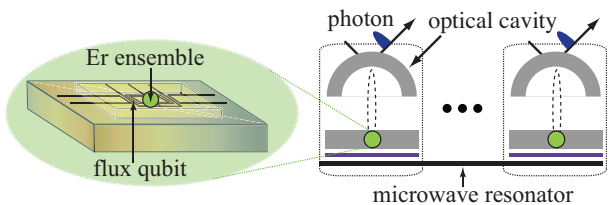}}
\end{center} 
\vspace*{-10pt}
\caption{A hybrid-system implementation of a repeater node, featuring erbium spins. Local gates are performed using gap-tunable flux qubits and a microwave resonator.}
\label{fig:implementation}
\end{figure}

We will outline how local gates are performed in the hybrid-system approach illustrated in Fig.~\ref{fig:implementation}. We will consider the situation where two ensembles in spatially separated cavities have information stored in them (they may be entangled with systems in remote nodes). Each ensemble encodes a qubit of information in the ground and first collectively-excited states. First and foremost, there is no direct coupling between the different ensembles so any two-qubit operation must be performed using a quantum bus-type approach \cite{qubus}, with flux qubits and a microwave resonator acting as the bus. 

Let us consider the individual components in turn and the couplings between them:

\begin{itemize}
\item {\it The ensemble}: An ensemble of $N$ Er$^{3+}$ electron spins can be described by the Hamiltonian
\begin{equation}
H=\frac{\hbar}{2} \sum _{k=1}^N g_e \mu_B B_z\; \sigma_z^{(k)},
\end{equation}
where $g_e=7$, $\mu_B$ is the Bohr magneton, $B_z$ is magnetic field in the $z$ axis, and $\sigma_z^{(k)}$ is the $z$ Pauli operator acting on the $k$th spin ($\sigma_\pm^{(k)}$ are the corresponding raising and lowering operators). We are considering the situation where there are few (zero or one) excitations in the ensemble, and so we can treat the ensemble of spins as a harmonic oscillator. The Hamiltonian can be rewritten as \cite{Marcos,Diniz}
\begin{equation}
H_{\rm ens}=\hbar \omega_{\rm ens} a^\dagger a,
\end{equation}
where $\hbar \omega_{\rm ens}$ is the energy splitting between levels caused the magnetic field and $a^\dagger$ and $a$ are the creation and destruction operators, defined as $a^\dagger=\frac{1}{\sqrt N}  \sum _{k=1}^N \sigma_+^{(k)},\;\;a=\frac{1}{\sqrt N}  \sum _{k=1}^N \sigma_-^{(k)}$.

\item {\it The flux qubit}: A gap-tunable flux qubit is composed of four Josephson junctions in two loops: a main loop and a gap control loop \cite{Paauw,zhu}. The main loop encloses three junctions: two identical junctions with Josephson energy $E_{\rm J}$ and one shared with the gap control loop with Josephson energy  $\alpha E_{\rm J}/2 $. The two junctions in the gap control loop are identical and form a dc-SQUID with effective energy $E^{\rm eff}_{\rm J}=\alpha cos(\pi {\it \Phi}_{\alpha} / {\it \Phi}_0)E_J$, where ${\it \Phi}_{\alpha}$ is the flux threading the gap control loop and ${\it \Phi}_{0}$ is the flux quantum respectively. When the effective magnetic flux ${\it \Phi}_{\rm qb}$ (${\it \Phi}_{\rm qb}={\it \Phi}_{\varepsilon}+{\it \Phi}_{\alpha} /2$)  threading the qubit is close to $(n+1/2) {\it \Phi}_{0}$, the qubit Hamiltonian can be written as
\begin{equation}
 H_{\rm qb}=\frac{\hbar}{2} \left[ {\it \Delta}(\Phi_\alpha) \sigma_z +{\it \varepsilon} ({\it \Phi}_{\rm qb}) \sigma_x \right],
\end{equation}
where $\sigma_x$  and $\sigma_z$ are the usual Pauli matrices, ${\it \Delta}(\Phi_\alpha)$ is the energy gap, and ${\varepsilon} ({\it \Phi}_{\rm qb})=2I_{\rm p}({\it \Phi}_{\rm qb}-{\it \Phi}_{\rm 0}/2)$ is the energy spacing of the two classical current states. The gap ${\it \Delta}(\Phi_\alpha)$  can be \emph{in situ} tuned on a nanosecond timescale while keeping ${\it \varepsilon} ({\it \Phi}_{\rm qb})=0$  \cite{zhu}.

\item {\it Coupling between a flux qubit and an ensemble}: The coupling between a flux qubit and an ensemble can be represented by \cite{Zhu1,Marcos}
\begin{equation}
H_{\rm qb-ens}=\hbar g_{\rm ens} \sigma_x \left(a+a^\dagger \right),
\end{equation}
where $g_{\rm ens}$ is the collective coupling constant. Considering the full Hamiltonian $H=H_{\rm qb}+H_{\rm ens}+H_{\rm qb-ens}$, we can move to a rotating frame, giving
\begin{equation}
H_{\rm rot-frame}=\hbar g_{\rm ens} \left (a^\dagger  \sigma_-+a \sigma_+\right),
\end{equation}
which is the well known Jaynes-Cummings interaction. This interaction leads to an $i\SWAP$ operation, which allows one to transfer the state of the flux qubit to the ensemble (or vice-versa) in a time $t_{\rm swap}=\pi/ g_{\rm ens}$. With $g_{\rm ens}/ 2\pi \sim 30$ MHz being possible \cite{couplingstrength}, the transfer time could be of the order of $t_{\rm swap} \sim 17$ ns.

This coupling also allows us to prepare the ensemble in an arbitrary superposition of its ground and first excited states, by creating the state in the flux qubit and then swapping it to the ensemble. 

\item {\it Coupling between a flux qubit and  a microwave resonator---direct and dispersive coupling}: A microwave resonator can generally be represented as a harmonic oscillator \cite{franco}
\begin{equation}
H_{\rm reson}=\hbar \omega_{\rm reson} b^\dagger b,
\end{equation}
where $\omega_{\rm reson}$ is the angular frequency of the oscillator and $b^\dagger$ and $b$ are the associated creation and destruction operators, respectively. The coupling between a flux qubit and a microwave resonator is generally given by 
\begin{equation}
H_{\rm qb-reson}=\hbar  g_{\rm qb-reson} \left (b^\dagger  \sigma_-+b\; \sigma_+\right),
\end{equation}
with coupling constant $g_{\rm qb-reson}$. We have two interesting regimes of operation: on-resonance and dispersive. Like before, the on-resonance coupling enables an $i\SWAP$ operation, this time to transfer the state of the flux qubit to the microwave resonator (or vice-versa) in a time $t_{\rm sw}= \pi / g_{\rm qb-reson}$. The dispersive regime has an effective Hamiltonian of the form 
\begin{equation}
 H_{\rm disp}=\frac{\hbar g_{\rm qb-reson}^2}{\delta} \;\; b^\dagger b\; \sigma_z,
\end{equation}
allowing a $\CZ$ (controlled-$Z$) gate to be performed, where $\delta$ is the detuning between the two systems. The $\CZ$ gate occurs at a time $t_{\rm disp} =\pi \delta /2 g_{\rm qb-reson}^2$, significantly longer than $t_{\rm sw}$. With $g_{\rm qb-reson}  / 2 \pi \sim 50$ MHz \cite{ref strength} and $\delta / 2 \pi \sim 500$ MHz we have $t_{\rm sw}\sim 10$ ns and $t_{\rm disp} \sim 50$ ns respectively.
\end{itemize}

The complete Hamiltonian for the two ensembles, the two flux qubits, and the microwave resonator is $H_{\rm total}=H_{\rm ens,1}+H_{\rm ens,2}+ H_{\rm qb,1}+ H_{\rm qb,2}+H_{\rm qb1-ens1}+H_{\rm qb2-ens2}+H_{\rm reson}+ H_{\rm qb1-reson}+ H_{\rm qb2-reson}$. We are now able to describe a two-qubit gate between ensemble qubits in the same node.

Consider the case where two ensembles have been prepared in the states $\vert a\rangle=a_0\vert0\rangle+a_1\vert1\rangle$ and $\vert b\rangle=b_0\vert0\rangle+b_1\vert1\rangle$, while the two gap-tunable flux qubits, $\vert f_1\rangle$ and $\vert f_2\rangle$, are initially prepared in the ground state, and the microwave resonator $\vert r\rangle$ is in a vacuum state. The flux qubits are off-resonant with both the ensemble qubits and the microwave resonator. Our initial state is
\begin{equation}
\vert a\rangle\vert f_1\rangle\vert r\rangle\vert f_2\rangle\vert b\rangle=\vert a\rangle\vert 0\rangle\vert 0\rangle\vert 0\rangle\vert b\rangle.
\end{equation}
The operation begins by bringing both flux qubits onto resonance with their associated ensembles, where the state of each ensemble can be transferred to its flux qubit using an $i\SWAP$ operation. This gives us the state 
\begin{equation}
\vert 0\rangle\vert a\rangle\vert 0\rangle\vert b\rangle\vert 0\rangle.
\end{equation}
Both flux qubits are then rapidly moved far off-resonance with the ensembles. The first flux qubit is brought onto resonance with the microwave resonator and an $i\SWAP$ operation is performed, giving us the state
\begin{equation}
\vert 0\rangle\vert 0\rangle\vert a\rangle\vert b\rangle\vert 0\rangle = \vert 0\rangle\vert 0\rangle\left[ a_0\vert0\rangle+a_1\vert1\rangle  \right] \left[  b_0\vert0\rangle+b_1\vert1\rangle \right] \vert 0\rangle.
\end{equation}
The flux qubit is then far detuned from the microwave resonator. The second flux qubit is brought into the dispersive limit with the microwave resonator, such that an interaction of the form $U = \exp \left[-i \frac{g_{\rm qb-reson}^2}{2 \delta}\;\;  t \;\; b^\dagger b\; \sigma_z \right]$ can be applied for a time $t_{\rm disp}$. The flux qubit is then moved far off-resonance with the microwave resonator, effectively decoupling the two systems. A simple single-qubit $\pi/2$ $Z$-rotation on the microwave resonator gives us the state
\begin{equation}
\vert 0\rangle\vert 0\rangle\left[ a_0 b_0 \vert0\rangle\vert0\rangle+ a_0 b_1 \vert0\rangle\vert1\rangle+ a_1 b_0 \vert1\rangle\vert0\rangle- a_1 b_1 \vert1\rangle\vert1\rangle \right] \vert 0\rangle.
\end{equation}
Finally, the quantum state stored in the microwave resonator is transferred to the first flux qubit (using the $i$SWAP operation) and the states of both flux qubits are transferred to their associated ensembles. The final state of the two ensembles is
\begin{equation}
 a_0 b_0 \vert0\rangle\vert0\rangle+ a_0 b_1 \vert0\rangle\vert1\rangle+ a_1 b_0 \vert1\rangle\vert0\rangle- a_1 b_1 \vert1\rangle\vert1\rangle.
\end{equation}
That is, an effective $\CZ$ gate has been performed between the two ensemble qubits. This operation can be completed in approximately 100 ns. For most purification protocols the $\CNOT$ gate is more useful than the $\CZ$ gate, but this can be achieved by applying a Hadamard gate to the second flux qubit before its state is transferred back to the ensemble.

We have described the protocol for performing local gates in a very general way. Whether these gates can be performed with the required fidelity remains an open question. 

\subsection{Coupling between the flux qubit and ensemble}

Coupling between a flux qubit and a single Er$^{3+}$ electron spin can be written in a rotating frame as
\begin{eqnarray}
H_{\rm int} = \hbar g  \left(\sigma_{\rm qb}^-  \sigma_{\rm er}^+ + \sigma_{\rm qb}^+  \sigma_{\rm er}^- \right),
\end{eqnarray}
where $\sigma_{\rm qb}^\pm$ are the raising and lower operators for the flux qubit, $\sigma_{\rm er}^\pm$ are the raising and lower operators for the Er$^{3+}$ atom, and $g$ is the coupling constant between them. This coupling constant can be roughly estimated as
\begin{eqnarray}
g/2\pi= g_e \mu_B B,
\end{eqnarray}
where $g_e = 7$ is the $g$ factor, $\mu_B = 14$ MHz/mT is the Bohr magneton, and $B$ is the magnetic field, which can be estimated using the Biot-Savart law. To first order, $B = \mu_0 I_p /(2R)$, where $\mu_0 = 4\pi \times  10^{-7}\; {\rm N  A}^{-2}$, $R=1/2\;\mu$m is the distance between the flux qubit and the single Er$^{3+}$ electron spin, and $I_p=1 \; \mu$A the persistent current in the qubit \cite{Zhu1,Marcos}. The coupling constant between the flux qubit and the electron spin can then be estimated as $g /2 \pi \sim 120$ kHz.

Now, for an ensemble of spins, our interaction term becomes 
\begin{eqnarray}
H_{\rm int} &=& \hbar \sum_k  g_k  \left(\sigma_{\rm qb}^- \sigma_{k,\rm er}^+ + \sigma_{\rm qb}^+  \sigma_{k,\rm er}^- \right) \nonumber \\
&=& \hbar g_{\rm ens} \left(\sigma_{\rm qb}^-  a^\dagger + \sigma_{\rm qb}^+ a \right),
\end{eqnarray}
where we have assumed that only a few excitations are in the ensemble at any one time, so that $a=\frac{1}{\sqrt{g_{\rm ens}}} \sum_k   g_k \sigma_{k,\rm er}^-$ with $g^2_{\rm ens}=\sum_k g_k^2$. If $g_k \sim g$ for all the electron spins, then $g_{\rm ens} =\sqrt{N} g$---that is, an ensemble of $N$ spins will give a $\sqrt N$ enhancement, allowing in principle a collective coupling constant of $g_{\rm ens}/ 2 \pi= 120 \sqrt{N}$ kHz. With $N=62$ $500$ we have $g_{\rm ens}/ 2 \pi= 30$ MHz. For an Er crystal of volume 40 $\mu {\rm m}^2 \times$ $1/2$ $\mu {\rm m}$ (with 40 $\mu {\rm m}^2$ corresponding to the area of the flux qubit \cite{Zhu1}), $N=62$ $500$ corresponds to an Er$^{3+}$ concentration of $1.5\times 10^{15}$ spins per cm$^3$. Such concentrations are readily commercially available \cite{endnote}. 

As our qubit is stored in the ensemble over a relatively long timescale (milliseconds) we need to examine its coherence properties. At 1.6 K the lifetime has been estimated at 100 ms \cite{Baldit1}. As our system contains superconducting circuits, we must operate at dilution fridge temperature (10 to 50 mK) and so we would expect the lifetime to be even longer. The $T_2$ coherence is more important, yet its value has not been determined at the concentrations being considered here. This coherence parameter will be affected by other impurities in the crystal, but at the concentrations proposed here, dipole-dipole interaction between Er$^{3+}$ ions should be negligible. Regardless, we require $T_2>10$  ms for our protocol to be effective.

\section{Discussion} 

As our scheme is based on fault-tolerant error correction, it is not surprising that the requirements on local gates and quantum memories are quite stringent. Superconducting circuits and quantum memories have not achieved this level of accuracy yet, but progress towards this goal has been made. Error rates of quantum gates in superconducting circuits are approaching values as low as 1\% \cite{Wallraff1}, and quantum memories are being engineered with increasing stability \cite{Steger1,Maurer1}. It may be possible to ease the requirements on these components by increasing the number of qubits per node. On the other hand, it would be interesting to study a scheme with only a few qubits per node, which may be much easier to implement. Such a scheme would involve more non-deterministic elements, thereby lowering the threshold \cite{Li2, Li3}.

Finally, our scheme enables universal quantum computation. In this context, node separation may be shorter and photon loss may be less severe. This may make the scheme an interesting avenue for the further study of fault-tolerant quantum computation in hybrid systems.

\acknowledgements{We thank X.~Zhu and S.~Saito for valuable discussions. This work was supported in part by JSPS, MEXT, FIRST, and NICT in Japan.}

\bibliographystyle{unsrt}

\begin{thebibliography}{10}
\bibitem{optimistic}
If we assume that $p_{single} = p_{coupling} = p_{detector}=0.75$, then $p_{\rm tot} \sim 0.2$.
\bibitem{sim}
We use a simulator written for this purpose and verified by reproducing the results of Ref.~\cite{Barrett1}. Our numerical results rely on the Blossom V matching algorithm \cite{Kolmogorov1} and the Mersenne Twister pseudorandom number generator \cite{Saito1}. Each data point in Fig.~\ref{fig:threshold} is an average of at least $10^5$ trials.
\bibitem{spacing note}
We assume that the rate of loss is the same for connections along all three axes of the cluster state. However, in practice, nodes might only be separated along the direction of communication, meaning that loss due to attenuation in the perpendicular directions will be significantly lower.
\end{thebibliography}

\end{document}